\documentclass[aip,jap,reprint,amsmath,amssymb,a4]{revtex4-1}
\usepackage{graphicx}
\usepackage{dcolumn}
\usepackage{bm}
\usepackage{braket}

\begin{document}


\title{First-principles investigation of polarization and ion conduction mechanisms in hydroxyapatite} 



\author{Shusuke Kasamatsu}
\email{kasamatsu@issp.u-tokyo.ac.jp}
\author{Osamu Sugino}%
\affiliation{%
The Institute for Solid State Physics, the University of Tokyo \\
5--1--5 Kashiwanoha, Kashiwa-shi, Chiba 277-8581, Japan 
}%

\date{\today}
\begin{abstract}
We report first-principles simulation of polarization mechanisms in hydroxyapatite to explain the underlying mechanism behind the reported ion conductivities and polarization under electrical poling at elevated temperatures. It is found that ion conduction occurs mainly in the column of OH$^-$ ions along the $c$-axis through a combination of the flipping of OH$^-$ ions, exchange of proton vacancies between OH$^-$ ions, and the hopping of the OH$^-$ vacancy. The calculated activation energies are consistent with those found in conductivity measurements and
thermally stimulated depolarization current measurements.

\end{abstract}


\maketitle


\section{Introduction}
Hydroxyapatite [Ca$_{10}$(PO$_4$)$_6$(OH)$_2$] (HAp) is a well-known mineral comprising up to 50\% of human bone, and many studies on this material have focused on
its use as a biomaterial for, e.g., bone transplants. HAp
is also known to exhibit ionic conductivity. Furthermore, nanocrystalline HAp has been shown to exhibit \mbox{piezo-,} pyro-, and ferroelectricity \cite{Lang2013}, opening up discussion for possible use of the material in various {\it in vivo} and {\it ex vivo} energy harvesting and nanoelectronic devices.


On a related but slightly different note, Tanaka and coworkers have reported that this material works remarkably well as an inorganic electret material \cite{Tanaka2010,Yamashita1996}. An electret exhibits a permanent dipole and can
be understood as
an analogue of a magnet with electrical polarization instead of magnetization. By applying DC bias at elevated temperatures and then quenching to room
temperature, hydroxyapatite can be poled into electrets. This is explained by the displacement of protons, whose motion at room temperature
is frozen due to the relatively high activation energy for diffusion. Thermally stimulated
depolarization current (TSDC) spectra shows that in polycrystalline samples, there are up to four polarization mechanisms with different activation energies and relaxation
times \cite{Tanaka2010, Horiuchi2012}.  Understanding of the microscopic mechanisms is highly desirable for optimizing this material for electret applications such as micro-energy harvesting and 
biological tissue engineering.

\begin{figure}[tb]
\centering
\includegraphics[width=\columnwidth]{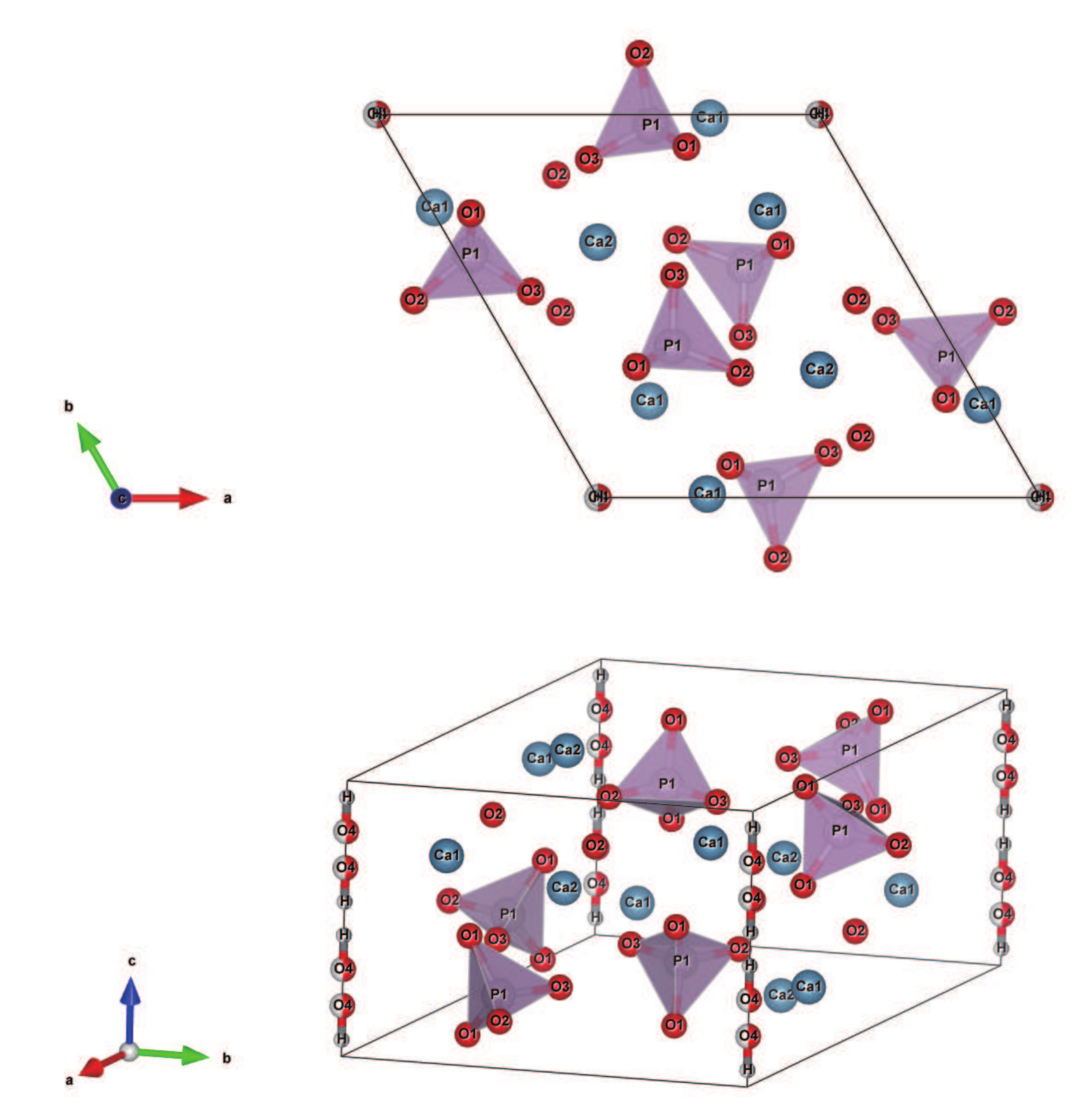}
\caption{\label{fig:HApStruct} The $P6_3/m$ crystal structure of hydroxyapatite.}
\end{figure}

Figure \ref{fig:HApStruct} shows the hexagonal $P6_3/m$ structure of HAp, which is the most frequently encountered structure, especially at elevated temperatures \cite{Ma2009, Yashima2014}. The crystal structure consists of columns of hydroxyl (OH$^-$ ) ions along the $c$-axis, PO$_4^{3-}$ tetrahedra, and Ca$^{2+}$ ions.  The occupancy of the O4 and H sites comprising the OH$^-$ ions is 1/2, which may be understood as originating from the orientational and occupational disorder of OH$^-$ ions at elevated temperature. That is, the structure along the $c$-axis can be understood as two OH sites centered at $z=1/4$ and $z=3/4$, and OH ions occupy each site in either the H-up or H-down orientation.
Our first-principles simulation results presented below confirm this view. At lower temperatures, there is some discrepancy as to the reported structures, probably owing to different stoichiometries, dopants, defects, disorder, and the intricacies involved in Rietveld fitting of diffraction data \cite{Elliott1973, Haverty2005, Ma2009, Lang2013}. Proposed structures include $P6_3$, $P2_1$, $P2_1/b$ and $P2_1/c$, which are all closely related to the $P6_3/m$ structure and differing in slight deviation from the hexagonal angles, slight canting of the PO$_4^{3-}$ tetrahedra, and the ordering of the OH$^-$ direction.

Previously, it was suggested that the ionic conductivity originates from proton hopping between the OH$^-$ ions and the PO$_4^{3-}$ tetrahedra along the $c$-axis based on the aging characteristics of the conductivity \cite{Yamashita1995}. Another work, based on electromotive force and electrolysis measurements on oxygen concentration cells, concluded that OH$^-$ ions are the main conducting species, not protons \cite{Takahashi1978}.
A more recent neutron diffraction work claims a sinusoidal proton diffusion pathway along the $c$-axis \cite{Yashima2014}, but neutron diffraction can only examine probability of finding protons at certain positions and is not a direct proof of long-range diffusion. Moreover, this can account for only one of the four polarization mechanisms noted above.
Atomistic simulations should be very useful for such investigation, although there are none, to our knowledge, that focus on ion diffusion in this material.
Thus, in this work, we turn to first-principles simulation to investigate possible microscopic mechanisms that can lead to polarization in HAp.
The minimum energy pathway and energy barriers for several hypothesized mechanisms are
calculated and their relation to various experiments are discussed.

It is well accepted experimentally and theoretically that HAp can accommodate a large number of defects in the OH$^-$ column that is introduced through dehydration \cite{Yamashita1995, Matsunaga2007}.
That is, one OH$^-$ ion and one proton are removed from the OH$^-$ column as a water molecule. This leaves one OH$^-$ vacancy (V$_{\text{OH}}^\bullet$) and one O$^{2-}$ ion (O$^\prime_\text{OH}$) at two lattice sites filled originally by OH$^-$ ions (Kr\"oger-Vink notation is given in parentheses). O$^\prime_\text{OH}$ can also be referred to as a proton vacancy (V$_\text{H}^\prime$); we use these terms interchangeably depending on whether we are discussing O$^{2-}$ diffusion or H$^+$ diffusion.

It is also known that Ca-deficient HAp can form depending on process conditions \cite{Tanaka2009}; in that case, it was shown through DFT calculations for 
defect formation energetics that proton interstitials can exist in 
substantial amounts to satisfy charge neutrality \cite{Matsunaga2008}. The interstitial protons are found to be most stable when forming a H$_2$O molecule-like 
structure by binding with OH$^-$ ions of the parent lattice. Therefore, we consider the influence of the OH$^-$ vacancy ($\text{V}^\bullet_\text{OH}$), proton vacancy (O$^\prime_\text{OH}$ or V$_\text{H}^\prime$), and proton interstitial on the polarization and ion conduction in this material.

\section{Method and model}
The calculations are performed using VASP \cite{Kresse1996,Kresse1996a} code based on the Kohn-Sham formalism of density functional theory (KS-DFT) \cite{Hohenberg1964,Kohn1965}. The GGA-PBE parametrization of the exchange-correlation functional \cite{Perdew1996, Perdew1996a} is employed. The projector-augmented wave (PAW) method \cite{Blochl1994} is used to describe ion-electron interactions, and the wave functions are expanded by a plane wave basis set with a cutoff energy of 500 eV for lattice-parameter relaxation, and 400 eV for the rest of the calculations. The structural relaxations are performed until forces on each ion become smaller than $10^{-2}$ eV/\r{A}. 
First-principles $NVT$ molecular dynamics simulations were carried out with a time step of 0.5 fs using the Nos\'e thermostat.
The minimum energy paths and energy barriers of several candidate polarization/diffusion mechanisms
were calculated using the climbing-image nudged elastic band (CI-NEB) method \cite{Henkelman2000a}.


In this work, we consider as the basis structure for our investigation the hexagonal unit cell shown in Fig.~\ref{fig:HApStruct}. Polarization and ion diffusion mechanisms are investigated using a supercell where the unit cell shown in Fig.~\ref{fig:HApStruct} is extended by a factor of three in the $c$-axis direction. We note that since we are considering diffusion of ionic species through defects at elevated temperatures, subtle differences in the lattice symmetry mentioned in Sec.~I are not very relevant to our work. We consider the possibility of intrinsic proton diffusion in the perfectly stoichiometric HAp. In addition, we consider defect-mediated diffusion of proton, hydroxyl, and oxide ions. When a single charged defect is present in the calculation cell, a uniform background charge is introduced to avoid the divergence of the electrostatic energy. Other defects may be examined in the future, although their contribution to the total ionic conductivity and poling under bias should be magnitudes smaller due to higher formation energies, as reported in Ref.~\onlinecite{Matsunaga2007, Matsunaga2008}.

\section{Results and discussion}
\begin{figure}[tb]
\centering
\includegraphics[width=0.8\columnwidth]{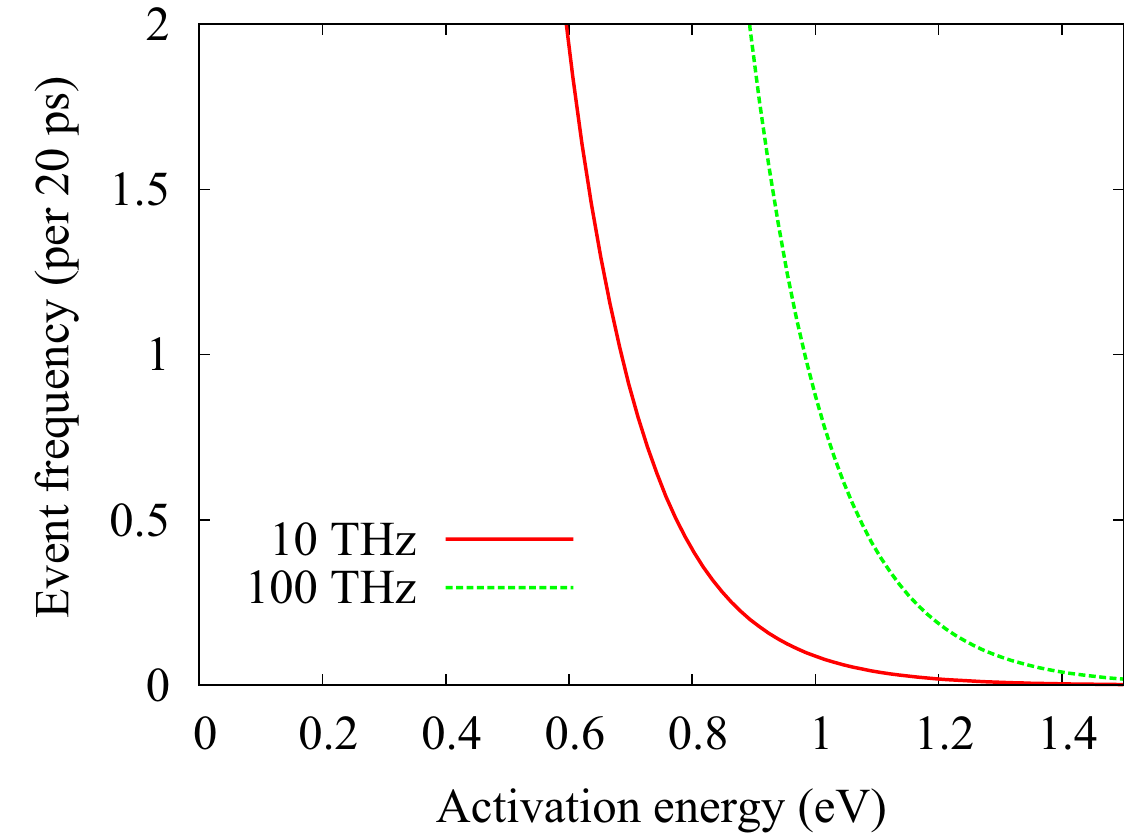}
\caption{\label{fig:Ea-gamma} The event frequency $\gamma = \nu \exp (- E_\text{a}/k_\text{B}T)$ at $T = 1500$ K for activated processes with a frequency prefactor $\nu$ of 10 THz (solid line) and 100 THz (dashed line) calculated as a function of the activation energy $E_\text{a}$ .}
\end{figure}

\begin{figure}[tb]
\centering
\includegraphics[width=\columnwidth]{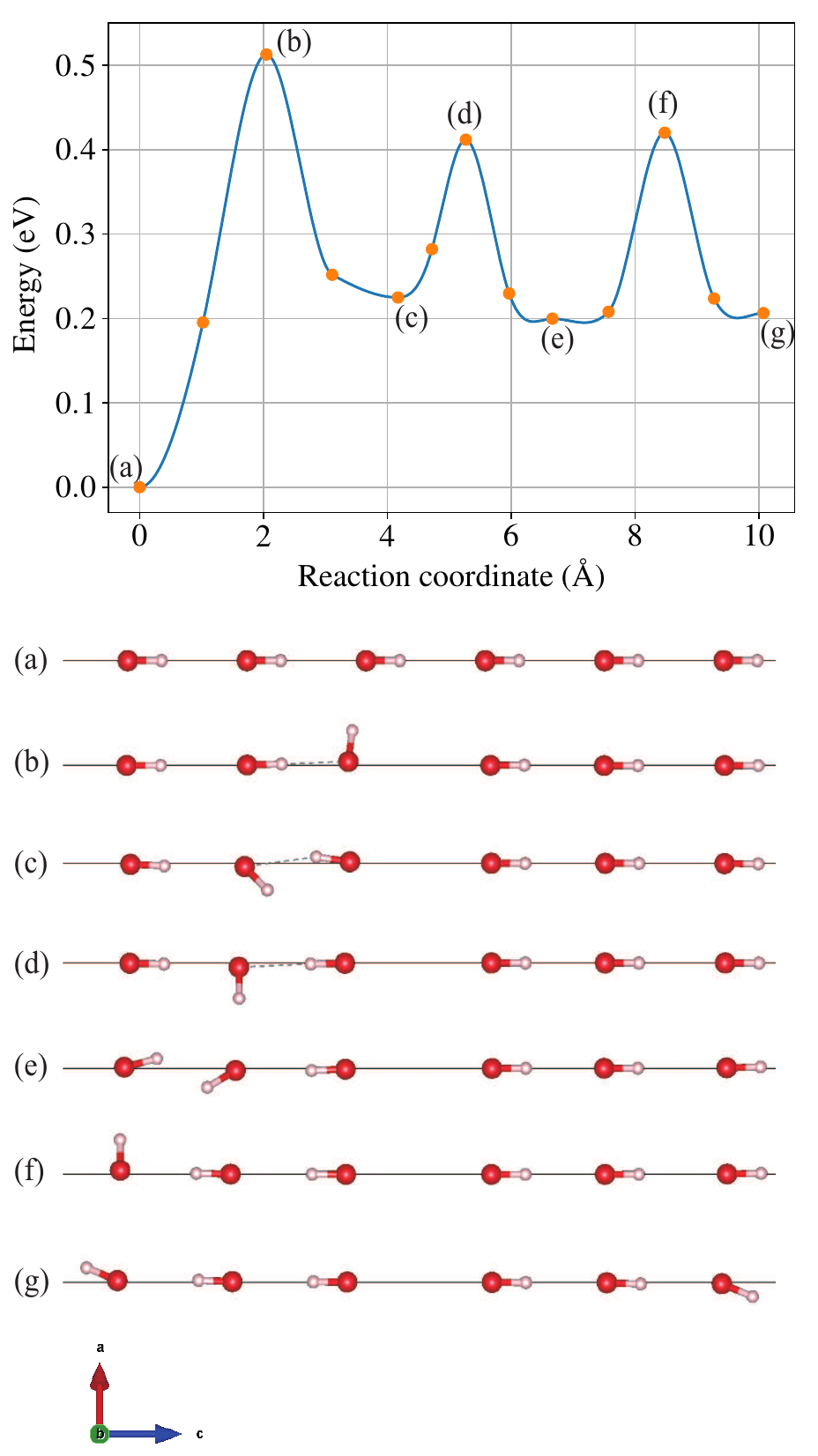}
\caption{\label{fig:OHflip} The minimum energy path for the flipping of OH$^-$ ions from a uniformly polarized configuration (a) to a half-polarized configuration (g). The structure of the OH$^-$ ions (a)--(g) along the minimum energy path are given at the bottom, while their energies along the reaction coordinate are given at the top. Filled circles correspond to the energies of the NEB images used in the calculation, while the lines between the filled circles are obtained through the interpolation scheme of Ref.~\onlinecite{Henkelman2000}.}
\end{figure}

\begin{table}[tb]
  \centering
  \caption{Comparison of the lattice parameter $a (=b)$, $c$, and  fractional coordinate  $z$ of the O and H atoms in the OH column of HAp calculated in this work with previous first-principles and experimental results.}
\begin{ruledtabular}
    \begin{tabular}{lcccc}
            & $a (=b)$ (\r{A}) & $c$ (\r{A}) & $z$ (O) & $z$ (H) \\
\hline
    This work (GGA-PBE)  & 9.552    & 6.912   & 0.276 & 0.418 \\
    GGA-PBE (Ref.~\onlinecite{Matsunaga2007}) & 9.536    & 6.904    & 0.286 & 0.428 \\
    LDA (Ref.~\onlinecite{Calderin2003}) & 9.11    & 6.86    & 0.29 & 0.42  \\
    Experiment (Ref.~\onlinecite{Kim2000})& 9.4302 & 6.8911 & 0.298 & 0.442 \\
    Experiment (Ref.~\onlinecite{Henning2001}) & 9.4214 & 6.8814 & -- & -- \\
    \end{tabular}%
\end{ruledtabular}
  \label{Tab:latcons}%
\end{table}%

As part of the initial setup for
subsequent calculations of diffusion barriers, a simultaneous relaxation of the lattice and internal parameters was performed on the unit cell of HAp (Fig.~\ref{fig:HApStruct}) where all OH$^-$ ions were aligned in the same direction. A $2 \times 2 \times 3$ k-point mesh centered at the $\Gamma$ point was employed for the Brillouin zone integration. Table \ref{Tab:latcons} compares the calculated lattice parameters with previous theoretical and experimental works. The result is within the typical error range for the GGA-PBE functional when compared to experiment, and is also in good agreement with previous theoretical works employing the same functional. In subsequent calculations, the unit cell was multiplied by a factor of three in the $c$-axis direction and the lattice parameters were kept fixed. We employed a smaller $2 \times 2 \times 1$ k-point mesh due to the extension of the supercell in the $c$-axis direction.

First, we performed molecular dynamics (MD) simulations at 1,000 -- 1,500 K for $>20$ ps as a quick-and-dirty way to find likely polarization and diffusion mechanisms. To estimate the possibility of observing activated processes within the MD simulation time, we calculated event frequencies as a function of activation energy $E_\text{a}$ based on harmonic transition state theory \cite{Vineyard1957} (Fig.~\ref{fig:Ea-gamma}). Considering the fact that characteristic frequencies of metal oxides are usually of the order of $\sim 10$ THz, and also considering the fact that we have hydrogen in the system, we can estimate the frequency prefactor $\nu$ entering the rate equation $\gamma = \nu \exp (- E_\text{a}/k_\text{B}T)$ to be of the order to 10--100 THz. Thus, from Fig.~\ref{fig:Ea-gamma}, we can expect to observe processes with $E_\text{a}$ of less than 0.6 eV many times within the simulation, while the chances of observing processes with $E_\text{a} > 1.2$ eV are nil. 

With the above in mind, we calculated systems with 1) no defects, 2) a proton interstitial, 3) a proton vacancy, and 4) proton vacancy--OH vacancy pair. Flipping of the direction of the OH$^-$ ions were observed in all cases, while proton interstitial hopping (HOH--OH $\leftrightarrow$ HO--HOH) in the OH column was observed in 2), OH--O $\leftrightarrow$ O--HO proton exchange was observed in 3) and 4), and OH$^-$ vacancy migration was observed in 4). That is, we were able to (luckily) detect migration processes for every defect considered. The fact that they were detected within such short simulations implies that they are also the dominant elementary processes for each ionic specie. Each of these mechanisms is examined in more detail below by using the CI-NEB method.

\subsection{Microscopic mechanisms for polarization and ion diffusion}
\subsubsection{OH flip}

As mentioned above, HAp has been shown to be ferroelectric, and the most natural mechanism for polarization reversal would be the flipping of the OH$^-$ ions in the $\pm c$-axis directions. Figure \ref{fig:OHflip} shows the minimum energy path for the reversal of the direction of one OH$^-$ ion starting from the uniformly polarized system, as well as subsequent flipping of neighboring OH$^-$ ions. We find that the flipping of the first OH$^-$ ion, which may be considered the nucleation of an opposite polarization domain in the 1D ferroelectric,  shows the highest activation energy at $\sim 0.5$ eV. The activation energy for subsequent enlargement of the domain is lower than 0.25 eV. It is interesting to note that when the hydrogen of the OH$^-$ ions face each other, the OH$^-$ ions are canted in the $a$ or $b$-axis directions to minimize the repulsion. Also, the total energies of the locally stable structures depend little on the ``domain size''; the energy is more or less constant when there are two ``domain walls'' in the unit cell at about 0.2 eV higher than the monodomain state. 
The relatively low activation energy combined with entropic effects means that the direction of the OH$^-$ ions would be highly disordered at elevated temperature. This may explain the 1/2 occupation of the O4 and H sites (see Fig.~\ref{fig:HApStruct}) along the OH$^-$ columns in the Rietveld fitting of the $P6_3/m$ structure. We also note that the flipping motion of the OH$^-$ ions are consistent with proton density mapping based on neutron diffraction measurements \cite{Yashima2014}.

\subsubsection{Formation of intrinsic defects}
In ceramic crystals, ions diffuse through crystals through defects such as vacancies and interstitials, which can be introduced in significant numbers through aliovalent ion doping or in the case of HAp, dehydration as noted above. Although much smaller in number, we may still consider the formation of intrinsic defects which are generated thermally within the material.
For example, a proton in one of the OH$^-$ ions may be detached and may bond with an adjacent OH$^-$ (OH--OH $\leftrightarrow$ O--HOH), creating what may be considered a proton vacancy-interstitial pair \cite{Gobinda1981}. 
Other authors have suggested that the PO$_4^{3-}$ tetrahedra around the OH$^-$ columns can trap protons \cite{Yamashita1995,Tanaka2010}.
We calculated the energy cost for the former reaction to be 1.9 eV and the latter as 1.8 eV, which are too high to explain the experimental results \cite{Yamashita1995,Tanaka2010},
where the highest activation energy reported is $\sim 1.3$ eV. The latter result is also consistent with a previous DFT calculation, which found that a structure where the O and H atoms of an OH$^-$ ion are split apart is unstable, and the atoms spontaneously recombine upon relaxation to form the OH$^-$ ion \cite{DeLeeuw2002}. In other words, splitting covalent O--H bonds costs too much energy without any help from electrostatics (i.e., charge neutrality condition) induced by aliovalent doping or dehydration. Thus, intrinsic defects generated through thermal splitting of OH$^-$ have virtually no contribution to the properties of this material. 

\begin{figure}[tb]
\centering
\includegraphics[width=\columnwidth]{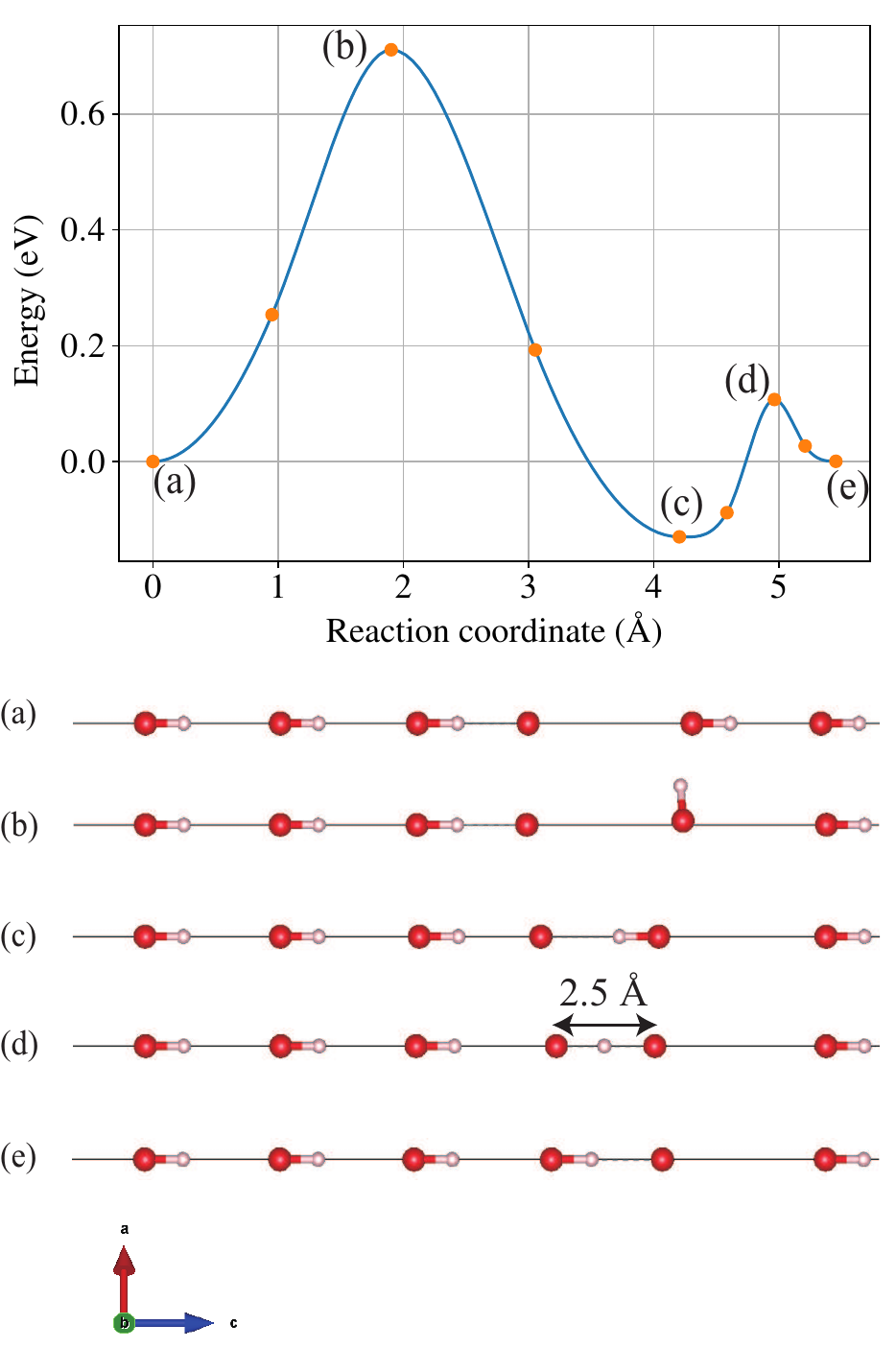}
\caption{\label{fig:Hjump} The minimum energy path for the migration of a proton vacancy. Flipping of an OH$^-$ ion next to a proton vacancy (a)--(c) and exchange of protons between adjacent oxide ions (c)--(e) must occur in series for long-range proton vacancy diffusion.}
\end{figure}

\begin{figure*}[tb]
\centering
\includegraphics[width=2\columnwidth]{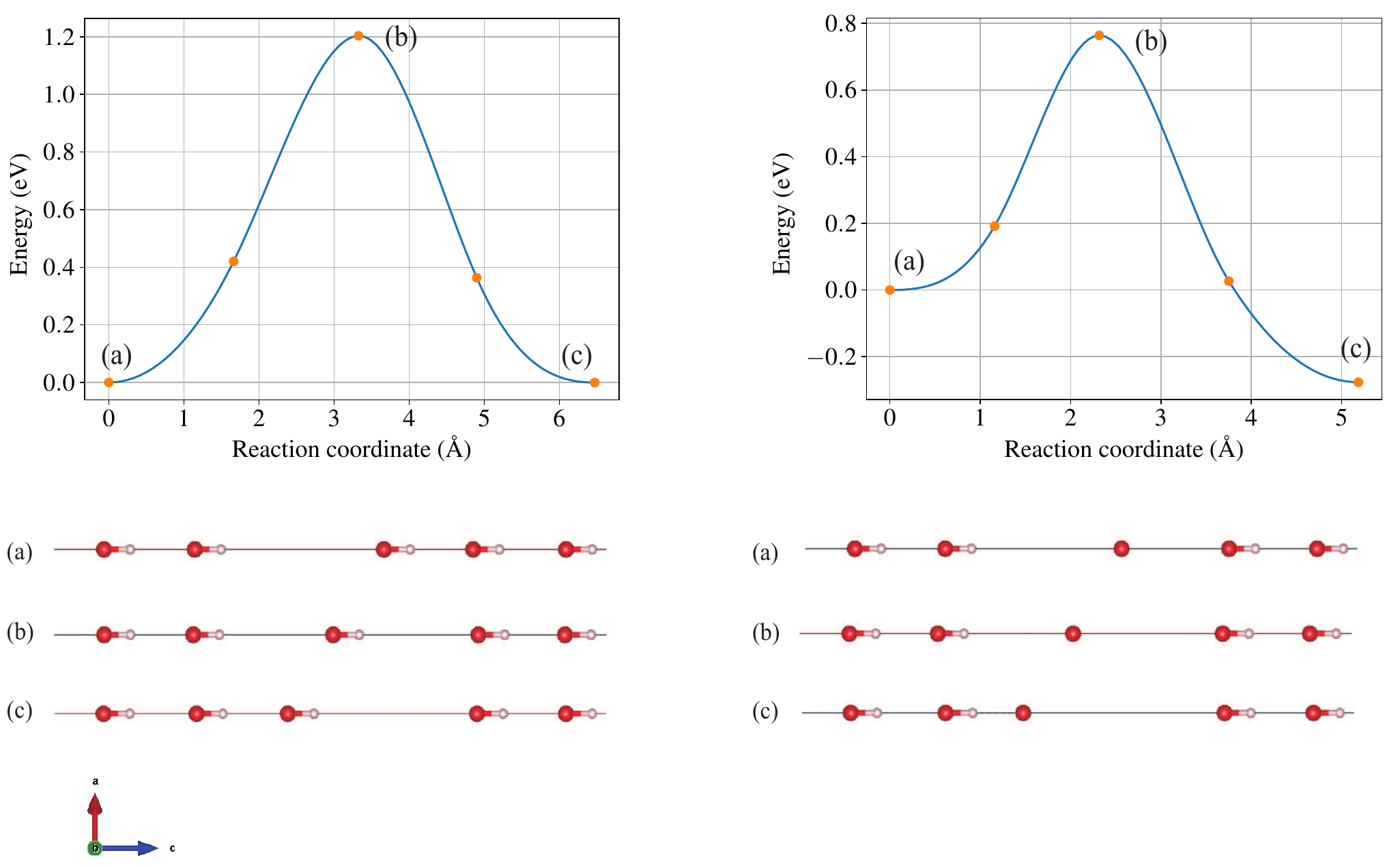}
\caption{\label{fig:OHjump} The minimum energy path for the migration of a OH$^-$ vacancy. The left figures depict $\text{V}_\text{OH}^\bullet$--$\text{OH}_\text{OH}^\text{x}$ exchange, while the right figures depict $\text{V}_\text{OH}^\bullet$--$\text{O}_\text{OH}^\prime$ exchange. }
\end{figure*}

\subsubsection{Proton vacancy}

Next, we consider the role of extrinsic defects introduced through dehydration or charge compensation for cation non-stoichiometries. 

First, we consider the migration of proton vacancies.
Figure \ref{fig:Hjump} shows one possible chain of events leading to proton vacancy diffusion: flipping of an OH$^-$ ion adjacent to a proton vacancy (--O--OH-- $\leftrightarrow$ --O--HO--), followed by exchange of protons between adjacent oxide ions 
(--O--HO-- $\leftrightarrow$ --OH--O--). This is basically the proton migration mechanism proposed in Ref.~\onlinecite{Yashima2014} based on neutron diffraction measurements.
In some previous works, proton hopping between OH$^-$ ions in the OH$^-$ column were deemed unviable because the distance between OH$^-$ ions are longer than typical O--O distances in hydrogen bond networks \cite{Takahashi1978, Yamashita1995}. However, we find that actually, the proton (vacancy) exchange
occurs rather easily with an activation energy of $\sim 0.2$ eV [Fig~\ref{fig:Hjump}(c)--(e)]. This is because as depicted in Fig.~\ref{fig:Hjump} (d), OH$^-$ ions can vibrate around its stable position without costing too much energy and the O--O distance at the transition state is only 2.5 \r{A}, which is certainly close enough for the shuttling of protons much like the Grotthus mechanism in the network of water molecules. On the other hand, it is notable that the OH$^-$ flipping next to a proton vacancy requires a higher activation energy compared to the pristine nondefective system examined in Fig.~\ref{fig:OHflip}.

Figs.~\ref{fig:Hjump} (a)--(e) is not the only possible chain of events leading to long range proton migration.  For example, the proton vacancy can exchange continuously from right to left without OH flipping as
\begin{equation*}
\text{--OH--OH--O--} \\
\rightarrow \text{--OH--O--HO--} \\
\rightarrow \text{--O--HO--HO--}
\end{equation*}
We stress, however, that macroscopic polarization involving the motion of more than one proton vacancy cannot occur without OH flipping. This is easily illustrated if we consider another proton vacancy coming from the right as --HO--HO--O--. The HO immediately to the left of O must first flip for the proton vacancy to move to the left as
\begin{equation*}
 \text{--HO--HO--O--} \rightarrow \text{--HO--OH--O--} \rightarrow \text{--HO--O--HO--}.
\end{equation*}

In dehydrated HAp, OH$^-$ vacancies exist in significant amounts in addition to proton vacancies, and this is expected to hinder the diffusion of protons \cite{Yamashita1995}. Indeed, we calculated the energy barrier for the traversal of a proton across an OH$^-$ vacancy (--O--V$_\text{OH}$--HO-- $\leftrightarrow$ --O--H--O-- $\leftrightarrow$ --OH--V$_\text{OH}$--O--)
to be higher than 4 eV, meaning that such direct proton diffusion across OH$^-$ vacancies is unlikely to occur in dehydrated HAp. Thus, we need to consider other mechanisms for long-range proton migration.

\subsubsection{OH$^-$ vacancy}
In the presence of OH$^-$ vacancies, it would be logical to consider the possibility of their migration through a vacancy hopping mechanism.
If the OH$^-$ vacancy hopping occurs at relatively low activation energies, then the protons do not need to traverse across OH$^-$ vacancies for long range diffusion. Instead, OH$^-$ ions can carry protons through vacancy hopping, or a combination of OH$^-$ vacancy hopping, OH$^-$ flipping, and \mbox{--OH--O-- $\leftrightarrow$ --O--HO--} proton exchange discussed in preceding sections would allow for long-range ion diffusion. We calculated the OH$^{-}$--$\text{V}_\text{OH}$ vacancy exchange energy to be $\sim 1.2$ eV (Fig.~\ref{fig:OHjump} left), and the OH$^-$--$\text{V}_\text{OH}$ vacancy exchange energy to be $\sim 0.8$--$1.1$ eV (Fig.~\ref{fig:OHjump} right, depending on the direction of diffusion), when all OH ions point in the same direction. The variation in the activation energies show that the precise values depend on the polarization of the environment surrounding the migrating ion.

\subsubsection{Proton interstitial}
\begin{figure*}[tb]
\centering
\includegraphics[width=2\columnwidth]{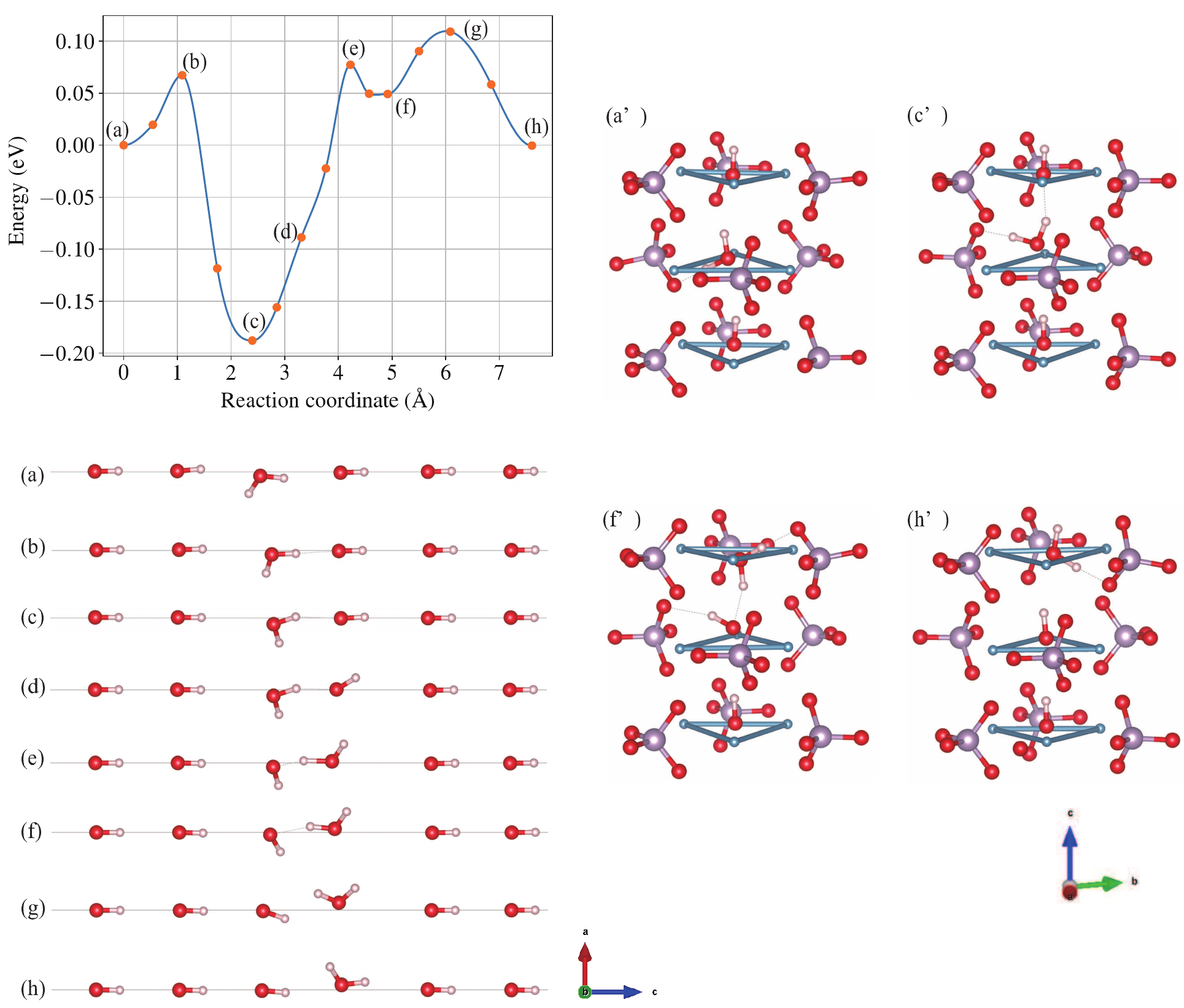}
\caption{\label{fig:Hinter} The minimum energy path for the migration of a proton interstitial within the OH$^-$ column. The energy along the path is given at the upper left, while the corresponding structures along the migration path is shown in the lower left and right-side figures.}
\end{figure*}
\begin{figure}[tb]
\centering
\includegraphics[width=\columnwidth]{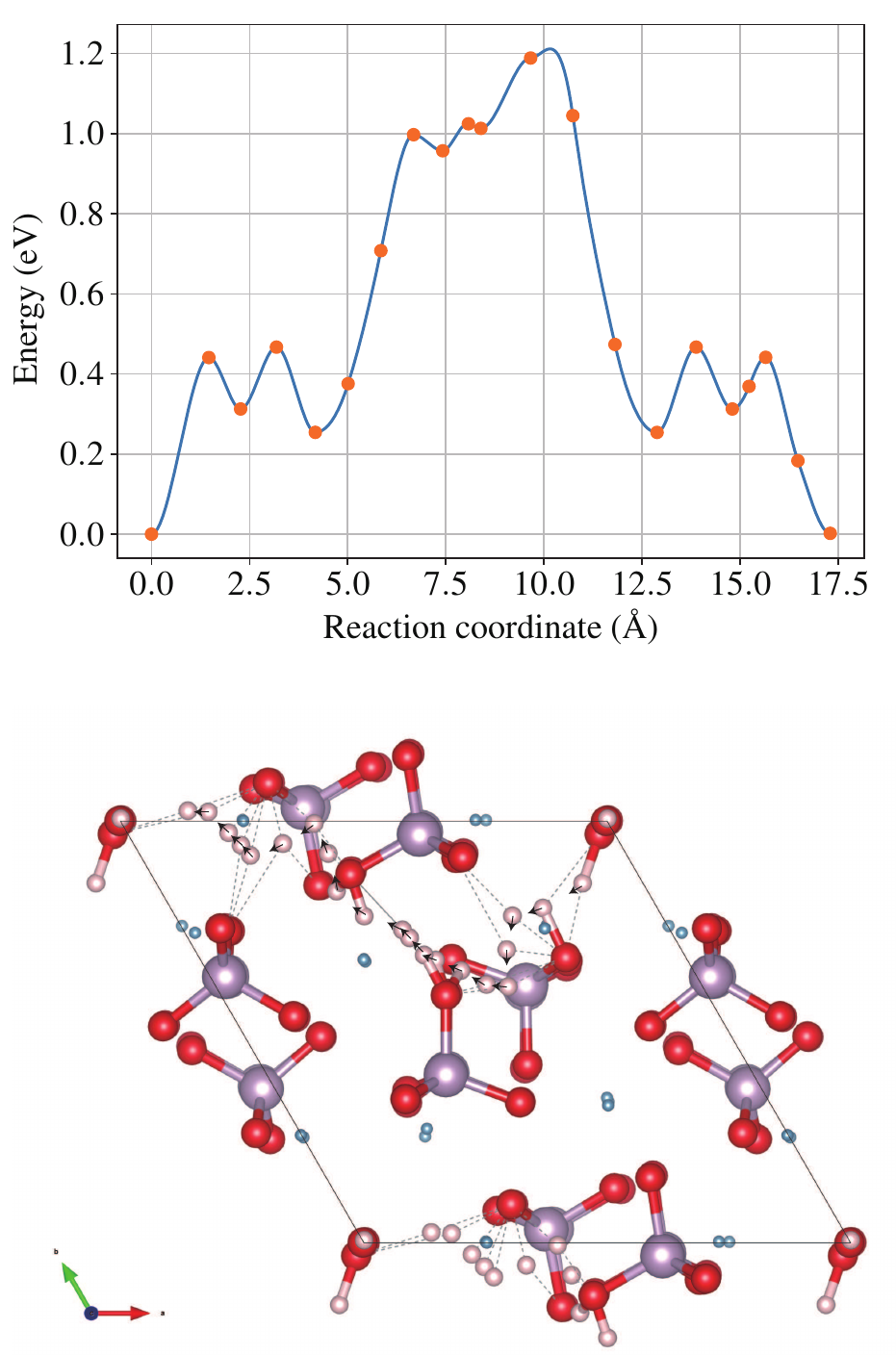}
\caption{\label{fig:Hinter_ab} The minimum energy path for the migration of the proton interstitial. The energy along the pathway is shown in the upper figure, while the migration of the proton interstitial is shown in the bottom. Only the motion of the migrating proton is shown while all other atoms are displayed at their initial positions. Arrows are guides for the eye. A full animation is provided in the online supplement.}
\end{figure}
In Ref.~\onlinecite{Matsunaga2008}, it was reported that
interstitial protons can be stabilized in low pH conditions in Ca-deficient HAp, and that the most stable position for the interstitial proton is on the OH$^-$ ion, forming an H$_2$O-molecule like structure. Bonding with PO$_4^{3-}$
with an O--H$\cdots$O linkage with an adjacent PO$_4^{3-}$ group was also found to be locally stable. These configurations were also realized in our first-principles molecular dynamics simulations mentioned above. Hopping along the OH$^-$ column was also observed in the molecular dynamics simulation, and the migration pathway for this mechanism calculated using CI-NEB is shown in Fig.~\ref{fig:Hinter}. The migration proceeds with rotation of H$_2$O and OH$^-$ combined with proton hopping. The total energy barrier is $\sim 0.3$ eV, so this is indeed a viable mechanism for proton conduction in Ca-deficient HAp. 
It can be seen from Figures \ref{fig:Hinter} (a')--(h') that hydrogen bonding with neighboring PO$_4^{3-}$ groups is utilized to stabilize the structures along the minimum energy path. Ref.~\onlinecite{Matsunaga2008} also showed the hydrogen bonding by examining the defect wave function.

Although we did not observe proton interstitial migration outside of the OH$^-$ column in the molecular dynamics simulation (perhaps due to the limited simulation time), we searched possible migration pathways in the $ab$ plane and found the pathway shown in Fig.~\ref{fig:Hinter_ab}. The energy barrier is $\sim 1.2$ eV, which is much higher than that within the OH$^-$ column. Thus, we can conclude that the interstitial migration is dominated by the OH$^-$ column. However, migration in the $ab$ plane may become relevant when bias voltage is applied perpendicular to the $c$-axis during electrical poling of a single crystal sample.

\subsection{Comparison of calculated activation energies with experiment}

\begin{table}[b]
  \centering
  \caption{Calculated activation energies $E_\text{a}$ of dominant polarization and ion migration mechanisms.}
\begin{ruledtabular}
    \begin{tabular}{lc}
           Mechanism & $E_\text{a}$ \\
\hline
    OH flipping  & 0.2 -- 0.8 eV \\
    OH--O $\leftrightarrow$ O--HO exchange & 0.1 -- 0.2 eV \\
    OH vacancy hopping & 0.8--1.2 eV \\
    Interstitial proton hopping & 0.025--0.25 eV \\
    \end{tabular}%
\end{ruledtabular}
  \label{Tab:Ea}%
\end{table}%
Table \ref{Tab:Ea} tabulates the mechanisms and corresponding activation energies calculated in preceding sections. The lowest activation energy for a single OH flip is 0.2 eV as shown in Fig.~\ref{fig:OHflip} (c--e) and (e--g), which corresponds to the growth of the 1D polarization domain by sequential flipping of adjacent OH ions. The highest calculated activation energy for OH flipping is 0.8 eV, which corresponds to the flip occurring next to a proton vacancy as in Fig.~\ref{fig:Hjump} (c--a). As for \mbox{OH--O $\leftrightarrow$ O--HO} exchange, the calculated values are 0.2 eV and 0.1 eV in Fig.~\ref{fig:Hjump} (c--e) and (e--c). OH vacancy hopping occurs with activation energies of 0.8--1.2 eV according to Fig.~\ref{fig:OHjump}, depending on whether the vacancy-ion exchange occurs with OH or O, and also on the polarization of the environment in the initial and final states.
In the case of interstitial proton hopping along the OH column, there are several local minima. The lowest and highest activation energies are 0.025 eV and 0.25 eV in Fig.~\ref{fig:Hinter} (f--c) and (c--f) corresponding to the two directions in HO--HOH $\leftrightarrow$ HOH--OH proton exchange.
It should be noted that the calculated values do not incorporate all possible variations due to the environment surrounding the local hopping/flipping site. Since long-range coulombic interactions are prevalent in this ionic material, it would be quite difficult to perform an exhaustive calculation taking into account all possible OH orientation and defect arrangements around the local hopping/flipping site. However, we can still estimate the effect of the environment from the limited number of varying defect arrangements taken into account in this work, and they suggest that the variations are smaller than few tenths of an eV as in Table \ref{Tab:Ea}.

The activation energies of the elementary processes calculated in this work do not correspond directly to those obtained from macroscopic conductivity measurements. To perform a direct comparison, one would have to perform e.g., kinetic Monte Carlo simulation to simulate the macroscopic behavior due to the competition and cooperation between microscopic elementary processes. However, microscopic activation energies often do correlate, to a certain degree, to those obtained from the Arrhenius plots of macroscopic measurements [they turn out to be equal according to random walk theory if there is only one mechanism and if the rate of that mechanism does not depend on the configuration (see, e.g., Ref.~\onlinecite{Shewmon1989})]. 
In the following, we attempt to interpret available experimental data based on Table \ref{Tab:Ea}.
We note that the degree of dehydration, and thus the ion carrier concentration, can change during conductivity or depolarization measurements if removal or addition of H$_2$O is not blocked. In this case, the dehydration energy would also contribute to the experimentally obtained activation energies. Such effects are usually negligible due to the use of blocking electrodes in electrical measurements if the temperature is kept below $\sim 1000$ K \cite{Yamashita1995}. Therefore, in the following discussion, we assume that generation of thermally activated carriers do not contribute to the measured activation energies.

The literature on conductivity measurements of stoichiometric or partially dehydrated HAp \cite{Takahashi1978, Yamashita1995,  Bouhaouss2001, Tanaka2009, Tanaka2010a} point to at least two rate-determining processes depending on the stoichiometry and degree of dehydration; there seems to be one process with an activation energy of $\sim 0.7$ eV, and another with $\gtrapprox 1$ eV that becomes more dominant as dehydration proceeds in HAp. By comparing to Table \ref{Tab:Ea}, the former process may be explained by proton conduction through OH$^-$ flipping combined with --OH--O-- $\leftrightarrow$ --O--HO-- proton exchange, where the highest activation barrier was calculated to be $\sim 0.8$ eV. The latter process likely kicks in when many OH$^-$ vacancies are introduced due to dehydration; in this case, OH$^-$ vacancy hopping with an activation energy of $\sim 1$ eV becomes the rate-determining step in ion conduction along the OH column.  Furthermore, TSDC experiments \cite{Tanaka2010, Horiuchi2012} report up to four polarization relaxation processes: one with an activation energy of $\sim 0.4$ eV, two with  $\sim 0.7$ eV, and another with $\sim 1$ eV. The latter two activation energies are again very close to those calculated in this work, suggesting that the OH$^-$ flipping and OH$^-$ vacancy hopping can be considered the rate-determining microscopic mechanisms for at least two of the reported polarization relaxation processes.

On the other hand, Ca-deficient HAp shows a rather complicated temperature dependence in the conductivity \cite{Tanaka2009} due to its instability against dehydration and decomposition to stoichiometric HAp and  $\beta$-tricalcium phosphate Ca$_3$[PO$_4$]$_2$. An abnormally high conductivity was reported for Ca-deficient HAp whiskers in the temperature range of \mbox{250--400 $^\circ$C} \cite{Tanaka2009}. Since proton interstitials are expected to exist in significant amounts in Ca-deficient HAp \cite{Matsunaga2008}, such high conductivity may be attributed to the interstitial proton hopping in the OH$^-$ column with a relatively low activation energy of $\sim 0.25$ eV. 

For further correlation with experimental data, we would have to consider the effect of the microstructure; for example, existence of trapping/detrapping mechanisms at grain boundaries and proton conduction through interfacial water have been suggested in the literature \cite{Tanaka2010}. Multiscale simulations are necessary for examining polarization mechanisms at longer time and length scales such as formation of space charges \cite{Kasamatsu2011, Kasamatsu2012} and their evolution under ac bias \cite{Pornprasertsuk2007}. Such undertakings are beyond the scope of this work, but should be considered for further understanding of this and similar materials systems to be used in various electrical/electrochemical devices.

\section{Conclusion}
In this work, we examined polarization and ion diffusion processes in hydroxyapatite using first-principles simulation. We found that the dominant polarization/diffusion mechanisms mainly occur in the column of OH$^-$ ions aligned along the $c$-axis, and that the main processes are the flipping of the direction of OH$^-$ ions, exchange of protons between oxide ions (--OH--O-- $\leftrightarrow$ --O--HO--), and the hopping of OH$^-$ vacancies. When proton interstitials are present as in the case of Ca-deficient HAp, interstitial proton hopping occurs mainly on the OH$^-$ column, although diffusion in the $ab$ plane is also viable under bias. The calculated activation energies are consistent with conductivity and TSDC measurements.

%
%


%

\begin{acknowledgments}
The authors thank Yumi Tanaka of the Tokyo University of Science for fruitful discussion.
The calculations were performed on the SGI ICE XA system at Institute for Solid State Physics, the University of Tokyo.
This work was supported by CREST, Japan Science and Technology Agency.
S. K. is also supported by Grant-in-Aid for Young Scientists (B) (No. 15K20953) by Japan Society for the
Promotion of Science.
Atomic structure figures were created using the visualization software VESTA \cite{Momma2008}.
\end{acknowledgments}

\bibliography{article}

\begin{thebibliography}{35}%
\makeatletter
\providecommand \@ifxundefined [1]{%
 \@ifx{#1\undefined}
}%
\providecommand \@ifnum [1]{%
 \ifnum #1\expandafter \@firstoftwo
 \else \expandafter \@secondoftwo
 \fi
}%
\providecommand \@ifx [1]{%
 \ifx #1\expandafter \@firstoftwo
 \else \expandafter \@secondoftwo
 \fi
}%
\providecommand \natexlab [1]{#1}%
\providecommand \enquote  [1]{``#1''}%
\providecommand \bibnamefont  [1]{#1}%
\providecommand \bibfnamefont [1]{#1}%
\providecommand \citenamefont [1]{#1}%
\providecommand \href@noop [0]{\@secondoftwo}%
\providecommand \href [0]{\begingroup \@sanitize@url \@href}%
\providecommand \@href[1]{\@@startlink{#1}\@@href}%
\providecommand \@@href[1]{\endgroup#1\@@endlink}%
\providecommand \@sanitize@url [0]{\catcode `\\12\catcode `\$12\catcode
  `\&12\catcode `\#12\catcode `\^12\catcode `\_12\catcode `\%12\relax}%
\providecommand \@@startlink[1]{}%
\providecommand \@@endlink[0]{}%
\providecommand \url  [0]{\begingroup\@sanitize@url \@url }%
\providecommand \@url [1]{\endgroup\@href {#1}{\urlprefix }}%
\providecommand \urlprefix  [0]{URL }%
\providecommand \Eprint [0]{\href }%
\providecommand \doibase [0]{http://dx.doi.org/}%
\providecommand \selectlanguage [0]{\@gobble}%
\providecommand \bibinfo  [0]{\@secondoftwo}%
\providecommand \bibfield  [0]{\@secondoftwo}%
\providecommand \translation [1]{[#1]}%
\providecommand \BibitemOpen [0]{}%
\providecommand \bibitemStop [0]{}%
\providecommand \bibitemNoStop [0]{.\EOS\space}%
\providecommand \EOS [0]{\spacefactor3000\relax}%
\providecommand \BibitemShut  [1]{\csname bibitem#1\endcsname}%
\let\auto@bib@innerbib\@empty
\bibitem [{\citenamefont {Lang}\ \emph {et~al.}(2013)\citenamefont {Lang},
  \citenamefont {Tofail}, \citenamefont {Kholkin}, \citenamefont
  {Wojta{\'{s}}}, \citenamefont {Gregor}, \citenamefont {Gandhi}, \citenamefont
  {Wang}, \citenamefont {Bauer}, \citenamefont {Krause},\ and\ \citenamefont
  {Plecenik}}]{Lang2013}%
  \BibitemOpen
  \bibfield  {author} {\bibinfo {author} {\bibfnamefont {S.~B.}\ \bibnamefont
  {Lang}}, \bibinfo {author} {\bibfnamefont {S.~A.~M.}\ \bibnamefont {Tofail}},
  \bibinfo {author} {\bibfnamefont {A.~L.}\ \bibnamefont {Kholkin}}, \bibinfo
  {author} {\bibfnamefont {M.}~\bibnamefont {Wojta{\'{s}}}}, \bibinfo {author}
  {\bibfnamefont {M.}~\bibnamefont {Gregor}}, \bibinfo {author} {\bibfnamefont
  {A.~A.}\ \bibnamefont {Gandhi}}, \bibinfo {author} {\bibfnamefont
  {Y.}~\bibnamefont {Wang}}, \bibinfo {author} {\bibfnamefont {S.}~\bibnamefont
  {Bauer}}, \bibinfo {author} {\bibfnamefont {M.}~\bibnamefont {Krause}}, \
  and\ \bibinfo {author} {\bibfnamefont {A.}~\bibnamefont {Plecenik}},\ }\href
  {\doibase 10.1038/srep02215} {\bibfield  {journal} {\bibinfo  {journal} {Sci.
  Rep.}\ }\textbf {\bibinfo {volume} {3}},\ \bibinfo {pages} {2215} (\bibinfo
  {year} {2013})}\BibitemShut {NoStop}%
\bibitem [{\citenamefont {Tanaka}\ \emph
  {et~al.}(2010{\natexlab{a}})\citenamefont {Tanaka}, \citenamefont {Iwasaki},
  \citenamefont {Nakamura}, \citenamefont {Nagai}, \citenamefont {Katayama},\
  and\ \citenamefont {Yamashita}}]{Tanaka2010}%
  \BibitemOpen
  \bibfield  {author} {\bibinfo {author} {\bibfnamefont {Y.}~\bibnamefont
  {Tanaka}}, \bibinfo {author} {\bibfnamefont {T.}~\bibnamefont {Iwasaki}},
  \bibinfo {author} {\bibfnamefont {M.}~\bibnamefont {Nakamura}}, \bibinfo
  {author} {\bibfnamefont {A.}~\bibnamefont {Nagai}}, \bibinfo {author}
  {\bibfnamefont {K.}~\bibnamefont {Katayama}}, \ and\ \bibinfo {author}
  {\bibfnamefont {K.}~\bibnamefont {Yamashita}},\ }\href {\doibase
  10.1063/1.3265429} {\bibfield  {journal} {\bibinfo  {journal} {J. Appl.
  Phys.}\ }\textbf {\bibinfo {volume} {107}},\ \bibinfo {pages} {014107}
  (\bibinfo {year} {2010}{\natexlab{a}})}\BibitemShut {NoStop}%
\bibitem [{\citenamefont {Yamashita}, \citenamefont {Oikawa},\ and\
  \citenamefont {Umegaki}(1996)}]{Yamashita1996}%
  \BibitemOpen
  \bibfield  {author} {\bibinfo {author} {\bibfnamefont {K.}~\bibnamefont
  {Yamashita}}, \bibinfo {author} {\bibfnamefont {N.}~\bibnamefont {Oikawa}}, \
  and\ \bibinfo {author} {\bibfnamefont {T.}~\bibnamefont {Umegaki}},\ }\href
  {\doibase DOI 10.1021/cm9602858} {\bibfield  {journal} {\bibinfo  {journal}
  {Chem. Mater.}\ }\textbf {\bibinfo {volume} {8}},\ \bibinfo {pages} {2697}
  (\bibinfo {year} {1996})}\BibitemShut {NoStop}%
\bibitem [{\citenamefont {Horiuchi}\ \emph {et~al.}(2012)\citenamefont
  {Horiuchi}, \citenamefont {Nakamura}, \citenamefont {Nagai}, \citenamefont
  {Katayama},\ and\ \citenamefont {Yamashita}}]{Horiuchi2012}%
  \BibitemOpen
  \bibfield  {author} {\bibinfo {author} {\bibfnamefont {N.}~\bibnamefont
  {Horiuchi}}, \bibinfo {author} {\bibfnamefont {M.}~\bibnamefont {Nakamura}},
  \bibinfo {author} {\bibfnamefont {A.}~\bibnamefont {Nagai}}, \bibinfo
  {author} {\bibfnamefont {K.}~\bibnamefont {Katayama}}, \ and\ \bibinfo
  {author} {\bibfnamefont {K.}~\bibnamefont {Yamashita}},\ }\href {\doibase
  10.1063/1.4754298} {\bibfield  {journal} {\bibinfo  {journal} {J. Appl.
  Phys.}\ }\textbf {\bibinfo {volume} {112}},\ \bibinfo {pages} {074901}
  (\bibinfo {year} {2012})}\BibitemShut {NoStop}%
\bibitem [{\citenamefont {Ma}\ and\ \citenamefont {Liu}(2009)}]{Ma2009}%
  \BibitemOpen
  \bibfield  {author} {\bibinfo {author} {\bibfnamefont {G.}~\bibnamefont
  {Ma}}\ and\ \bibinfo {author} {\bibfnamefont {X.~Y.}\ \bibnamefont {Liu}},\
  }\href {\doibase 10.1021/cg900156w} {\bibfield  {journal} {\bibinfo
  {journal} {Cryst. Growth Des.}\ }\textbf {\bibinfo {volume} {9}},\ \bibinfo
  {pages} {2991} (\bibinfo {year} {2009})}\BibitemShut {NoStop}%
\bibitem [{\citenamefont {Yashima}\ \emph {et~al.}(2014)\citenamefont
  {Yashima}, \citenamefont {Kubo}, \citenamefont {Omoto}, \citenamefont
  {Fujimori}, \citenamefont {Fujii},\ and\ \citenamefont
  {Ohoyama}}]{Yashima2014}%
  \BibitemOpen
  \bibfield  {author} {\bibinfo {author} {\bibfnamefont {M.}~\bibnamefont
  {Yashima}}, \bibinfo {author} {\bibfnamefont {N.}~\bibnamefont {Kubo}},
  \bibinfo {author} {\bibfnamefont {K.}~\bibnamefont {Omoto}}, \bibinfo
  {author} {\bibfnamefont {H.}~\bibnamefont {Fujimori}}, \bibinfo {author}
  {\bibfnamefont {K.}~\bibnamefont {Fujii}}, \ and\ \bibinfo {author}
  {\bibfnamefont {K.}~\bibnamefont {Ohoyama}},\ }\href {\doibase
  10.1021/jp412771f} {\bibfield  {journal} {\bibinfo  {journal} {J. Phys. Chem.
  C}\ }\textbf {\bibinfo {volume} {118}},\ \bibinfo {pages} {5180} (\bibinfo
  {year} {2014})}\BibitemShut {NoStop}%
\bibitem [{\citenamefont {Elliott}, \citenamefont {Mackie},\ and\ \citenamefont
  {Young}(1973)}]{Elliott1973}%
  \BibitemOpen
  \bibfield  {author} {\bibinfo {author} {\bibfnamefont {J.~C.}\ \bibnamefont
  {Elliott}}, \bibinfo {author} {\bibfnamefont {P.~E.}\ \bibnamefont {Mackie}},
  \ and\ \bibinfo {author} {\bibfnamefont {R.~A.}\ \bibnamefont {Young}},\
  }\href {\doibase 10.1126/science.180.4090.1055} {\bibfield  {journal}
  {\bibinfo  {journal} {Science}\ }\textbf {\bibinfo {volume} {180}},\ \bibinfo
  {pages} {1055} (\bibinfo {year} {1973})}\BibitemShut {NoStop}%
\bibitem [{\citenamefont {Haverty}\ \emph {et~al.}(2005)\citenamefont
  {Haverty}, \citenamefont {Tofail}, \citenamefont {Stanton},\ and\
  \citenamefont {McMonagle}}]{Haverty2005}%
  \BibitemOpen
  \bibfield  {author} {\bibinfo {author} {\bibfnamefont {D.}~\bibnamefont
  {Haverty}}, \bibinfo {author} {\bibfnamefont {S.~A.~M.}\ \bibnamefont
  {Tofail}}, \bibinfo {author} {\bibfnamefont {K.~T.}\ \bibnamefont {Stanton}},
  \ and\ \bibinfo {author} {\bibfnamefont {J.~B.}\ \bibnamefont {McMonagle}},\
  }\href {\doibase 10.1103/PhysRevB.71.094103} {\bibfield  {journal} {\bibinfo
  {journal} {Phys. Rev. B}\ }\textbf {\bibinfo {volume} {71}},\ \bibinfo
  {pages} {094103} (\bibinfo {year} {2005})}\BibitemShut {NoStop}%
\bibitem [{\citenamefont {Yamashita}(1995)}]{Yamashita1995}%
  \BibitemOpen
  \bibfield  {author} {\bibinfo {author} {\bibfnamefont {K.}~\bibnamefont
  {Yamashita}},\ }\href {\doibase 10.1111/j.1151-2916.1995.tb08468.x}
  {\bibfield  {journal} {\bibinfo  {journal} {J. Am. Ceram. Soc.}\ }\textbf
  {\bibinfo {volume} {78}},\ \bibinfo {pages} {1191} (\bibinfo {year}
  {1995})}\BibitemShut {NoStop}%
\bibitem [{\citenamefont {Takahashi}, \citenamefont {Tanase},\ and\
  \citenamefont {Yamamoto}(1978)}]{Takahashi1978}%
  \BibitemOpen
  \bibfield  {author} {\bibinfo {author} {\bibfnamefont {T.}~\bibnamefont
  {Takahashi}}, \bibinfo {author} {\bibfnamefont {S.}~\bibnamefont {Tanase}}, \
  and\ \bibinfo {author} {\bibfnamefont {O.}~\bibnamefont {Yamamoto}},\
  }\href@noop {} {\bibfield  {journal} {\bibinfo  {journal} {Electrochim.
  Acta}\ }\textbf {\bibinfo {volume} {23}},\ \bibinfo {pages} {369} (\bibinfo
  {year} {1978})}\BibitemShut {NoStop}%
\bibitem [{\citenamefont {Matsunaga}\ and\ \citenamefont
  {Kuwabara}(2007)}]{Matsunaga2007}%
  \BibitemOpen
  \bibfield  {author} {\bibinfo {author} {\bibfnamefont {K.}~\bibnamefont
  {Matsunaga}}\ and\ \bibinfo {author} {\bibfnamefont {A.}~\bibnamefont
  {Kuwabara}},\ }\href {\doibase 10.1103/PhysRevB.75.014102} {\bibfield
  {journal} {\bibinfo  {journal} {Phys. Rev. B}\ }\textbf {\bibinfo {volume}
  {75}},\ \bibinfo {pages} {014102} (\bibinfo {year} {2007})}\BibitemShut
  {NoStop}%
\bibitem [{\citenamefont {Tanaka}\ \emph {et~al.}(2009)\citenamefont {Tanaka},
  \citenamefont {Nakamura}, \citenamefont {Nagai}, \citenamefont {Toyama},\
  and\ \citenamefont {Yamashita}}]{Tanaka2009}%
  \BibitemOpen
  \bibfield  {author} {\bibinfo {author} {\bibfnamefont {Y.}~\bibnamefont
  {Tanaka}}, \bibinfo {author} {\bibfnamefont {M.}~\bibnamefont {Nakamura}},
  \bibinfo {author} {\bibfnamefont {A.}~\bibnamefont {Nagai}}, \bibinfo
  {author} {\bibfnamefont {T.}~\bibnamefont {Toyama}}, \ and\ \bibinfo {author}
  {\bibfnamefont {K.}~\bibnamefont {Yamashita}},\ }\href {\doibase
  10.1016/j.mseb.2009.01.016} {\bibfield  {journal} {\bibinfo  {journal}
  {Mater. Sci. Eng., B}\ }\textbf {\bibinfo {volume} {161}},\ \bibinfo {pages}
  {115} (\bibinfo {year} {2009})}\BibitemShut {NoStop}%
\bibitem [{\citenamefont {Matsunaga}(2008)}]{Matsunaga2008}%
  \BibitemOpen
  \bibfield  {author} {\bibinfo {author} {\bibfnamefont {K.}~\bibnamefont
  {Matsunaga}},\ }\href {\doibase 10.1103/PhysRevB.77.104106} {\bibfield
  {journal} {\bibinfo  {journal} {Phys. Rev. B}\ }\textbf {\bibinfo {volume}
  {77}},\ \bibinfo {pages} {104106} (\bibinfo {year} {2008})}\BibitemShut
  {NoStop}%
\bibitem [{\citenamefont {Kresse}\ and\ \citenamefont
  {Furthm\"{u}ller}(1996{\natexlab{a}})}]{Kresse1996}%
  \BibitemOpen
  \bibfield  {author} {\bibinfo {author} {\bibfnamefont {G.}~\bibnamefont
  {Kresse}}\ and\ \bibinfo {author} {\bibfnamefont {J.}~\bibnamefont
  {Furthm\"{u}ller}},\ }\href@noop {} {\bibfield  {journal} {\bibinfo
  {journal} {Phys. Rev. B}\ }\textbf {\bibinfo {volume} {54}},\ \bibinfo
  {pages} {11169} (\bibinfo {year} {1996}{\natexlab{a}})}\BibitemShut {NoStop}%
\bibitem [{\citenamefont {Kresse}\ and\ \citenamefont
  {Furthm\"{u}ller}(1996{\natexlab{b}})}]{Kresse1996a}%
  \BibitemOpen
  \bibfield  {author} {\bibinfo {author} {\bibfnamefont {G.}~\bibnamefont
  {Kresse}}\ and\ \bibinfo {author} {\bibfnamefont {J.}~\bibnamefont
  {Furthm\"{u}ller}},\ }\href@noop {} {\bibfield  {journal} {\bibinfo
  {journal} {Comp. Mater. Sci.}\ }\textbf {\bibinfo {volume} {6}},\ \bibinfo
  {pages} {15} (\bibinfo {year} {1996}{\natexlab{b}})}\BibitemShut {NoStop}%
\bibitem [{\citenamefont {Hohenberg}\ and\ \citenamefont
  {Kohn}(1964)}]{Hohenberg1964}%
  \BibitemOpen
  \bibfield  {author} {\bibinfo {author} {\bibfnamefont {P.}~\bibnamefont
  {Hohenberg}}\ and\ \bibinfo {author} {\bibfnamefont {W.}~\bibnamefont
  {Kohn}},\ }\href@noop {} {\bibfield  {journal} {\bibinfo  {journal} {Phys.
  Rev.}\ }\textbf {\bibinfo {volume} {136}},\ \bibinfo {pages} {B864} (\bibinfo
  {year} {1964})}\BibitemShut {NoStop}%
\bibitem [{\citenamefont {Kohn}\ and\ \citenamefont {Sham}(1965)}]{Kohn1965}%
  \BibitemOpen
  \bibfield  {author} {\bibinfo {author} {\bibfnamefont {W.}~\bibnamefont
  {Kohn}}\ and\ \bibinfo {author} {\bibfnamefont {L.~J.}\ \bibnamefont
  {Sham}},\ }\href@noop {} {\bibfield  {journal} {\bibinfo  {journal} {Phys.
  Rev.}\ }\textbf {\bibinfo {volume} {140}},\ \bibinfo {pages} {A1133}
  (\bibinfo {year} {1965})}\BibitemShut {NoStop}%
\bibitem [{\citenamefont {Perdew}, \citenamefont {Burke},\ and\ \citenamefont
  {Ernzerhof}(1996)}]{Perdew1996}%
  \BibitemOpen
  \bibfield  {author} {\bibinfo {author} {\bibfnamefont {J.~P.}\ \bibnamefont
  {Perdew}}, \bibinfo {author} {\bibfnamefont {K.}~\bibnamefont {Burke}}, \
  and\ \bibinfo {author} {\bibfnamefont {M.}~\bibnamefont {Ernzerhof}},\ }\href
  {\doibase 10.1103/PhysRevLett.77.3865} {\bibfield  {journal} {\bibinfo
  {journal} {Phys. Rev. Lett.}\ }\textbf {\bibinfo {volume} {77}},\ \bibinfo
  {pages} {3865} (\bibinfo {year} {1996})}\BibitemShut {NoStop}%
\bibitem [{\citenamefont {Perdew}, \citenamefont {Burke},\ and\ \citenamefont
  {Ernzerhof}(1997)}]{Perdew1996a}%
  \BibitemOpen
  \bibfield  {author} {\bibinfo {author} {\bibfnamefont {J.~P.}\ \bibnamefont
  {Perdew}}, \bibinfo {author} {\bibfnamefont {K.}~\bibnamefont {Burke}}, \
  and\ \bibinfo {author} {\bibfnamefont {M.}~\bibnamefont {Ernzerhof}},\ }\href
  {\doibase 10.1103/PhysRevLett.78.1396} {\bibfield  {journal} {\bibinfo
  {journal} {Phys. Rev. Lett.}\ }\textbf {\bibinfo {volume} {78}},\ \bibinfo
  {pages} {1396} (\bibinfo {year} {1997})}\BibitemShut {NoStop}%
\bibitem [{\citenamefont {Bl\"{o}chl}(1994)}]{Blochl1994}%
  \BibitemOpen
  \bibfield  {author} {\bibinfo {author} {\bibfnamefont {P.~E.}\ \bibnamefont
  {Bl\"{o}chl}},\ }\href@noop {} {\bibfield  {journal} {\bibinfo  {journal}
  {Phys. Rev. B}\ }\textbf {\bibinfo {volume} {50}},\ \bibinfo {pages} {17953}
  (\bibinfo {year} {1994})}\BibitemShut {NoStop}%
\bibitem [{\citenamefont {Henkelman}, \citenamefont {Uberuaga},\ and\
  \citenamefont {Jónsson}(2000)}]{Henkelman2000a}%
  \BibitemOpen
  \bibfield  {author} {\bibinfo {author} {\bibfnamefont {G.}~\bibnamefont
  {Henkelman}}, \bibinfo {author} {\bibfnamefont {B.~P.}\ \bibnamefont
  {Uberuaga}}, \ and\ \bibinfo {author} {\bibfnamefont {H.}~\bibnamefont
  {Jónsson}},\ }\href {\doibase 10.1063/1.1329672} {\bibfield  {journal}
  {\bibinfo  {journal} {J. Chem. Phys.}\ }\textbf {\bibinfo {volume} {113}},\
  \bibinfo {pages} {9901} (\bibinfo {year} {2000})}\BibitemShut {NoStop}%
\bibitem [{\citenamefont {Henkelman}\ and\ \citenamefont
  {J{\'{o}}nsson}(2000)}]{Henkelman2000}%
  \BibitemOpen
  \bibfield  {author} {\bibinfo {author} {\bibfnamefont {G.}~\bibnamefont
  {Henkelman}}\ and\ \bibinfo {author} {\bibfnamefont {H.}~\bibnamefont
  {J{\'{o}}nsson}},\ }\href {\doibase 10.1063/1.1323224} {\bibfield  {journal}
  {\bibinfo  {journal} {J. Chem. Phys.}\ }\textbf {\bibinfo {volume} {113}},\
  \bibinfo {pages} {9978} (\bibinfo {year} {2000})}\BibitemShut {NoStop}%
\bibitem [{\citenamefont {Calder{\'{i}}n}, \citenamefont {Stott},\ and\
  \citenamefont {Rubio}(2003)}]{Calderin2003}%
  \BibitemOpen
  \bibfield  {author} {\bibinfo {author} {\bibfnamefont {L.}~\bibnamefont
  {Calder{\'{i}}n}}, \bibinfo {author} {\bibfnamefont {M.~J.}\ \bibnamefont
  {Stott}}, \ and\ \bibinfo {author} {\bibfnamefont {A.}~\bibnamefont
  {Rubio}},\ }\href {\doibase 10.1103/PhysRevB.67.134106} {\bibfield  {journal}
  {\bibinfo  {journal} {Phys. Rev. B}\ }\textbf {\bibinfo {volume} {67}},\
  \bibinfo {pages} {134106} (\bibinfo {year} {2003})}\BibitemShut {NoStop}%
\bibitem [{\citenamefont {Kim}\ \emph {et~al.}(2000)\citenamefont {Kim},
  \citenamefont {Fenton}, \citenamefont {Hunter},\ and\ \citenamefont
  {Kennedy}}]{Kim2000}%
  \BibitemOpen
  \bibfield  {author} {\bibinfo {author} {\bibfnamefont {J.~Y.}\ \bibnamefont
  {Kim}}, \bibinfo {author} {\bibfnamefont {R.~R.}\ \bibnamefont {Fenton}},
  \bibinfo {author} {\bibfnamefont {B.~A.}\ \bibnamefont {Hunter}}, \ and\
  \bibinfo {author} {\bibfnamefont {B.~J.}\ \bibnamefont {Kennedy}},\ }\href
  {\doibase http://dx.doi.org/10.1071/CH00060} {\bibfield  {journal} {\bibinfo
  {journal} {Aust. J. Chem.}\ }\textbf {\bibinfo {volume} {53}},\ \bibinfo
  {pages} {679} (\bibinfo {year} {2000})}\BibitemShut {NoStop}%
\bibitem [{\citenamefont {Alberius-Henning}\ \emph {et~al.}(2001)\citenamefont
  {Alberius-Henning}, \citenamefont {Adolfsson}, \citenamefont {Grins},\ and\
  \citenamefont {Fitch}}]{Henning2001}%
  \BibitemOpen
  \bibfield  {author} {\bibinfo {author} {\bibfnamefont {P.}~\bibnamefont
  {Alberius-Henning}}, \bibinfo {author} {\bibfnamefont {E.}~\bibnamefont
  {Adolfsson}}, \bibinfo {author} {\bibfnamefont {J.}~\bibnamefont {Grins}}, \
  and\ \bibinfo {author} {\bibfnamefont {A.}~\bibnamefont {Fitch}},\ }\href
  {\doibase 10.1023/A:1004876622105} {\bibfield  {journal} {\bibinfo  {journal}
  {J. Mater. Sci.}\ }\textbf {\bibinfo {volume} {36}},\ \bibinfo {pages} {663}
  (\bibinfo {year} {2001})}\BibitemShut {NoStop}%
\bibitem [{\citenamefont {Vineyard}(1957)}]{Vineyard1957}%
  \BibitemOpen
  \bibfield  {author} {\bibinfo {author} {\bibfnamefont {G.~H.}\ \bibnamefont
  {Vineyard}},\ }\href {\doibase 10.1016/0022-3697(57)90059-8} {\bibfield
  {journal} {\bibinfo  {journal} {J. Phys. Chem. Solids}\ }\textbf {\bibinfo
  {volume} {3}},\ \bibinfo {pages} {121} (\bibinfo {year} {1957})}\BibitemShut
  {NoStop}%
\bibitem [{\citenamefont {Gobinda}\ and\ \citenamefont
  {Freund}(1981)}]{Gobinda1981}%
  \BibitemOpen
  \bibfield  {author} {\bibinfo {author} {\bibfnamefont {C.~M.}\ \bibnamefont
  {Gobinda}}\ and\ \bibinfo {author} {\bibfnamefont {F.}~\bibnamefont
  {Freund}},\ }\href@noop {} {\bibfield  {journal} {\bibinfo  {journal} {J.
  Chem. Soc., Dalton Trans.}\ }\textbf {\bibinfo {volume} {0}},\ \bibinfo
  {pages} {949} (\bibinfo {year} {1981})}\BibitemShut {NoStop}%
\bibitem [{\citenamefont {de~Leeuw}(2002)}]{DeLeeuw2002}%
  \BibitemOpen
  \bibfield  {author} {\bibinfo {author} {\bibfnamefont {N.}~\bibnamefont
  {de~Leeuw}},\ }\href {\doibase 10.1039/b203114k} {\bibfield  {journal}
  {\bibinfo  {journal} {Phys. Chem. Chem. Phys.,}\ }\textbf {\bibinfo {volume}
  {4}},\ \bibinfo {pages} {3865} (\bibinfo {year} {2002})}\BibitemShut
  {NoStop}%
\bibitem [{\citenamefont {Shewmon}(1989)}]{Shewmon1989}%
  \BibitemOpen
  \bibfield  {author} {\bibinfo {author} {\bibfnamefont {P.}~\bibnamefont
  {Shewmon}},\ }\href@noop {} {\emph {\bibinfo {title} {Diffusion in solids}}}\
  (\bibinfo  {publisher} {The Minerals, Metals, \& Materials Society},\
  \bibinfo {year} {1989})\BibitemShut {NoStop}%
\bibitem [{\citenamefont {Bouhaouss}\ \emph {et~al.}(2001)\citenamefont
  {Bouhaouss}, \citenamefont {Laghzizil}, \citenamefont {Bensaoud},
  \citenamefont {Ferhat}, \citenamefont {Lorent},\ and\ \citenamefont
  {Livage}}]{Bouhaouss2001}%
  \BibitemOpen
  \bibfield  {author} {\bibinfo {author} {\bibfnamefont {A.}~\bibnamefont
  {Bouhaouss}}, \bibinfo {author} {\bibfnamefont {A.}~\bibnamefont
  {Laghzizil}}, \bibinfo {author} {\bibfnamefont {A.}~\bibnamefont {Bensaoud}},
  \bibinfo {author} {\bibfnamefont {M.}~\bibnamefont {Ferhat}}, \bibinfo
  {author} {\bibfnamefont {G.}~\bibnamefont {Lorent}}, \ and\ \bibinfo {author}
  {\bibfnamefont {J.}~\bibnamefont {Livage}},\ }\href {\doibase
  10.1016/S1466-6049(01)00054-X} {\bibfield  {journal} {\bibinfo  {journal}
  {Int. J. Inorg. Mater.}\ }\textbf {\bibinfo {volume} {3}},\ \bibinfo {pages}
  {743} (\bibinfo {year} {2001})}\BibitemShut {NoStop}%
\bibitem [{\citenamefont {Tanaka}\ \emph
  {et~al.}(2010{\natexlab{b}})\citenamefont {Tanaka}, \citenamefont {Kikuchi},
  \citenamefont {Tanaka}, \citenamefont {Hashimoto}, \citenamefont {Hojo},
  \citenamefont {Nakamura}, \citenamefont {Nagai}, \citenamefont {Sugiyama},
  \citenamefont {Munakata},\ and\ \citenamefont {Yamashita}}]{Tanaka2010a}%
  \BibitemOpen
  \bibfield  {author} {\bibinfo {author} {\bibfnamefont {Y.}~\bibnamefont
  {Tanaka}}, \bibinfo {author} {\bibfnamefont {M.}~\bibnamefont {Kikuchi}},
  \bibinfo {author} {\bibfnamefont {K.}~\bibnamefont {Tanaka}}, \bibinfo
  {author} {\bibfnamefont {K.}~\bibnamefont {Hashimoto}}, \bibinfo {author}
  {\bibfnamefont {J.}~\bibnamefont {Hojo}}, \bibinfo {author} {\bibfnamefont
  {M.}~\bibnamefont {Nakamura}}, \bibinfo {author} {\bibfnamefont
  {A.}~\bibnamefont {Nagai}}, \bibinfo {author} {\bibfnamefont
  {T.}~\bibnamefont {Sugiyama}}, \bibinfo {author} {\bibfnamefont
  {F.}~\bibnamefont {Munakata}}, \ and\ \bibinfo {author} {\bibfnamefont
  {K.}~\bibnamefont {Yamashita}},\ }\href {\doibase
  10.1111/j.1551-2916.2010.04105.x} {\bibfield  {journal} {\bibinfo  {journal}
  {J. Am. Ceram. Soc.}\ }\textbf {\bibinfo {volume} {93}},\ \bibinfo {pages}
  {3577} (\bibinfo {year} {2010}{\natexlab{b}})}\BibitemShut {NoStop}%
\bibitem [{\citenamefont {Kasamatsu}, \citenamefont {Tada},\ and\ \citenamefont
  {Watanabe}(2011)}]{Kasamatsu2011}%
  \BibitemOpen
  \bibfield  {author} {\bibinfo {author} {\bibfnamefont {S.}~\bibnamefont
  {Kasamatsu}}, \bibinfo {author} {\bibfnamefont {T.}~\bibnamefont {Tada}}, \
  and\ \bibinfo {author} {\bibfnamefont {S.}~\bibnamefont {Watanabe}},\ }\href
  {\doibase 10.1016/j.ssi.2010.11.022} {\bibfield  {journal} {\bibinfo
  {journal} {Solid State Ionics}\ }\textbf {\bibinfo {volume} {183}},\ \bibinfo
  {pages} {20} (\bibinfo {year} {2011})}\BibitemShut {NoStop}%
\bibitem [{\citenamefont {Kasamatsu}, \citenamefont {Tada},\ and\ \citenamefont
  {Watanabe}(2012)}]{Kasamatsu2012}%
  \BibitemOpen
  \bibfield  {author} {\bibinfo {author} {\bibfnamefont {S.}~\bibnamefont
  {Kasamatsu}}, \bibinfo {author} {\bibfnamefont {T.}~\bibnamefont {Tada}}, \
  and\ \bibinfo {author} {\bibfnamefont {S.}~\bibnamefont {Watanabe}},\ }\href
  {\doibase 10.1016/j.ssi.2012.08.009} {\bibfield  {journal} {\bibinfo
  {journal} {Solid State Ionics}\ }\textbf {\bibinfo {volume} {226}},\ \bibinfo
  {pages} {62} (\bibinfo {year} {2012})}\BibitemShut {NoStop}%
\bibitem [{\citenamefont {Pornprasertsuk}\ \emph {et~al.}(2007)\citenamefont
  {Pornprasertsuk}, \citenamefont {Cheng}, \citenamefont {Huang},\ and\
  \citenamefont {Prinz}}]{Pornprasertsuk2007}%
  \BibitemOpen
  \bibfield  {author} {\bibinfo {author} {\bibfnamefont {R.}~\bibnamefont
  {Pornprasertsuk}}, \bibinfo {author} {\bibfnamefont {J.}~\bibnamefont
  {Cheng}}, \bibinfo {author} {\bibfnamefont {H.}~\bibnamefont {Huang}}, \ and\
  \bibinfo {author} {\bibfnamefont {F.~B.}\ \bibnamefont {Prinz}},\ }\href
  {\doibase 10.1016/j.ssi.2006.12.016} {\bibfield  {journal} {\bibinfo
  {journal} {Solid State Ionics}\ }\textbf {\bibinfo {volume} {178}},\ \bibinfo
  {pages} {195} (\bibinfo {year} {2007})}\BibitemShut {NoStop}%
\bibitem [{\citenamefont {Momma}\ and\ \citenamefont
  {Izumi}(2008)}]{Momma2008}%
  \BibitemOpen
  \bibfield  {author} {\bibinfo {author} {\bibfnamefont {K.}~\bibnamefont
  {Momma}}\ and\ \bibinfo {author} {\bibfnamefont {F.}~\bibnamefont {Izumi}},\
  }\href@noop {} {\bibfield  {journal} {\bibinfo  {journal} {J. Appl.
  Crystallogr.}\ }\textbf {\bibinfo {volume} {41}},\ \bibinfo {pages} {653}
  (\bibinfo {year} {2008})}\BibitemShut {NoStop}%
\end{thebibliography}%

\end{document}